\documentclass[%
 reprint,
 amsmath,amssymb,
 aps,
]{revtex4-2}

\usepackage{color}
\usepackage{graphicx}
\usepackage{dcolumn}
\usepackage{bm}

\begin{document}

\preprint{APS/123-QED}

\title{Exact solutions of the simplified March model for organizational learning}

\author{Hang-Hyun Jo}
\email{h2jo@catholic.ac.kr}
\affiliation{Department of Physics, The Catholic University of Korea, Bucheon 14662, Republic of Korea}
\date{\today}

\begin{abstract}
James G. March's celebrated agent-based simulation model for organizational learning [March, Organization Science \textbf{2}, 71 (1991)] has been extensively studied for the last decades. Yet the model was not fully understood due to the lack of analytical solutions of the model. We simplify the March model to take an analytical approach using master equations. We then derive exact solutions for some simplest yet nontrivial cases, and perform numerical estimation of master equations for more complicated cases. Both analytical and numerical results are in good agreement with agent-based simulations. These results are also compared to those of the original March model. Our approach enables us to rigorously understand the results of the simplified model as well as the original model to a large extent.
\end{abstract}

\maketitle


\section{Introduction}

James G. March introduced an agent-based simulation model for organizational learning in his seminal paper in 1991~\cite{March1991Exploration}. Since then, the original March model, its variants, and other similar models have been extensively studied for the last decades~\cite{Rodan2005Exploration, Blaschke2006Forgotten, Miller2006Adding, Kane2007Information, Kim2009Exploration, Fang2010Balancing, Sachdeva2013Encounter, Schilling2014When, Chanda2015Optimal, Miller2016Organizational, Chanda2017Inferring, Chanda2018Continuum, Chanda2019Replicating, Marin-Idarraga2022Factors}. March's model considers an external reality, an organizational code (code hereafter), and individual members of the organization. The code represents a set of norms, rules, etc. that is updated using the knowledge of individuals about the reality, while individuals learn from the code about the reality too. By doing so, the organizational knowledge about the reality is collected from individuals and it is disseminated to them at the same time. He studied the effects of learning rates of individuals and of the code on their achieved knowledge about the reality. In order to consider more realistic situations, he took personnel turnover and environmental turbulence into account in his model to find that there may exist an optimal turnover rate that maximizes the achieved knowledge, depending on the learning rates. 

We remark that the March model can be considered in the framework of opinion dynamics in networks~\cite{Castellano2009Statistical, Acemoglu2011Opinion, Sen2014Sociophysics, Sirbu2017Opinion, Proskurnikov2017Tutorial, Proskurnikov2018Tutorial, Baronchelli2018Emergence, Anderson2019Recent, Noorazar2020Recent}. That is, the code plays a hub node in a hub-and-spoke network, while individuals are dangling nodes~\cite{Barabasi2016Network}. Nodes update their opinions or beliefs according to their neighbors' opinions or beliefs, while the external reality plays an external field or source affecting all nodes. It implies that various analytical approaches developed for the opinion dynamics can be applied to the March model.

The March model and its variants have provided with insights into management and business administration~\cite{Sachdeva2013Encounter} but mostly by means of computer simulations of models~\cite{Chanda2019Replicating}. In general, for rigorous understanding of models, derivation of their exact, analytical solutions is of utmost importance. In our work, we simplify March's original model to explicitly write down master equations describing the dynamics of the model. Then we derive exact solutions of the simplified model for some simplest yet nontrivial cases. Numerical estimation of master equations is performed for more complicated cases. Both analytical and numerical results are shown to be in good agreement with agent-based simulation results. Our approach enables us to rigorously understand the results of not only the simplified model but also the original model to a large extent.

The paper is organized as follows. In Sec.~\ref{sec:model}, we describe the original March model and our simplified version. In Sec.~\ref{sec:result}, analytical, numerical, and simulation results of the simplified model are presented and compared to the results of the original March model. Finally, we conclude our work in Sec.~\ref{sec:conclusion}.

\section{Models}\label{sec:model}

\subsection{Original March model}

As mentioned, the original model by March considers the external reality, the code, and individual members of the organization~\cite{March1991Exploration}. We remark that in this Subsection we will use mathematical symbols originally used in the March's paper, and they are not supposed to be confused with symbols in the next Subsection and throughout the paper. The model is based on the following assumptions.

(i) The reality is characterized in terms of an $m$-dimensional vector, each element of which may have the value of $1$ or $-1$ with equal probabilities.

(ii) The code and $n$ individuals in the organization have beliefs about the reality. The belief is also represented by an $m$-dimensional vector, each element of which may have the value of $1$, $0$, or $-1$ with equal probabilities. These beliefs may change over time.

(iii) At each time step, each individual may change elements of its belief that are different from those of the code unless the code's element is $0$. Each of such elements of the individual changes to that of the code with a probability $p_1$, independently of other elements.

(iv) At the same time, the code updates its belief based on beliefs of some individuals. For this, individuals whose beliefs are closer to the reality than the code is to the reality are identified, which are called a superior group. Then each element of the code's belief changes to the dominant element within the superior group with a probability $p_2$, independently of other elements.

So far, the reality has been assumed to be fixed and the individuals are not replaced by new ones. Thus, it is called a closed system. March first considered a homogeneous population in the closed system, in which all individuals are assigned the identical learning probability. Then he considered the heterogeneous population in the closed system, such that some individuals have higher learning probability than the others. Finally, a homogeneous population in the open system is also considered; in the open system individuals may be replaced by new ones (turnover) and/or the reality changes over time (turbulence). The turnover probability is denoted by $p_3$ and the turbulence probability is by $p_4$. That is, with a probability $p_3$ each individual is replaced by a new one having a random belief vector at each time step. Also, each element of the reality shifts to the other value, i.e., from $1$ to $-1$ or from $-1$ to $1$, with a probability $p_4$.

\subsection{Simplified March model}

Let us simplify the March model. As in the original model we consider an external reality, a code, and $N$ agents. At a time step $t$, the external reality, denoted by a variable $r(t)$, can have a value of $0$ or $1$. Beliefs of the code and agents about reality are respectively represented by variables, namely, $c(t)\in\{0,1\}$ for the code and $\sigma_i(t)\in\{0,1\}$ for the $i$th agent with $i=1,\ldots,N$. For a given initial condition of $r(0)$, $c(0)$, and $\{\sigma_i(0)\}$, each time step consists of four stages:

(i) Every agent of $i\in\{1,\ldots,N\}$ independently updates their belief by learning from the code with a \emph{socialization} probability $p^{(i)}\in (0,1]$:
\begin{align}
    \sigma_i(t+1)=c(t).
    \label{eq:agent_learn}
\end{align}

(ii) Each agent is replaced by a new agent with a \emph{turnover} probability $u\in[0,1]$, and the new agent is assigned a belief randomly drawn from $\{0,1\}$. 

(iii) The code learns from agents who are superior to the code. Here superior agents indicate those whose beliefs are closer to the reality than the code's belief is. For example, if $r(t)=1$, the code learns from superior agents only when $c(t)=0$ and there is at least one agent with $\sigma_i(t)=1$. Denoting a superior agent to the code by $j$, the code updates its belief with a \emph{codification} probability $q\in (0,1]$:
\begin{align}
    c(t+1)=\sigma_j(t)\ \textrm{for}\ j\in\{i|\delta_{\sigma_i(t), r(t)}>\delta_{c(t), r(t)}\},
    \label{eq:code_learn}
\end{align}
where $\delta_{\cdot,\cdot}$ is a Kronecker delta. 

(iv) With a \emph{turbulence} probability $v\in[0,1]$, the reality is assigned a new value randomly drawn from $\{0,1\}$, which closes the time step. 

Since the reality, the code, and agents update their value or beliefs synchronously, the order of four stages does not affect the result, except for the socialization and turnover of agents. Note that parameters $\{p^{(i)}\}$, $q$, $u$, and $v$ in our simplified model correspond to $p_1$, $p_2$, $p_3$, and $p_4$ in the original March model~\cite{March1991Exploration}, respectively.

Our simplified model can be called a closed system if $u=v=0$, otherwise it is an open system. In the open system, the reality can vary over time (turbulence), and/or agents can be replaced by new agents (turnover). In contrast, in the closed system, the reality is assumed to have the fixed value of $1$ for the entire period of time, i.e., $r(t)=1$ for all $t$, without loss of generality, and there is no turnover of agents.

\section{Results}\label{sec:result}

\begin{figure*}[!t]
\centering
\includegraphics[width=0.75\textwidth]{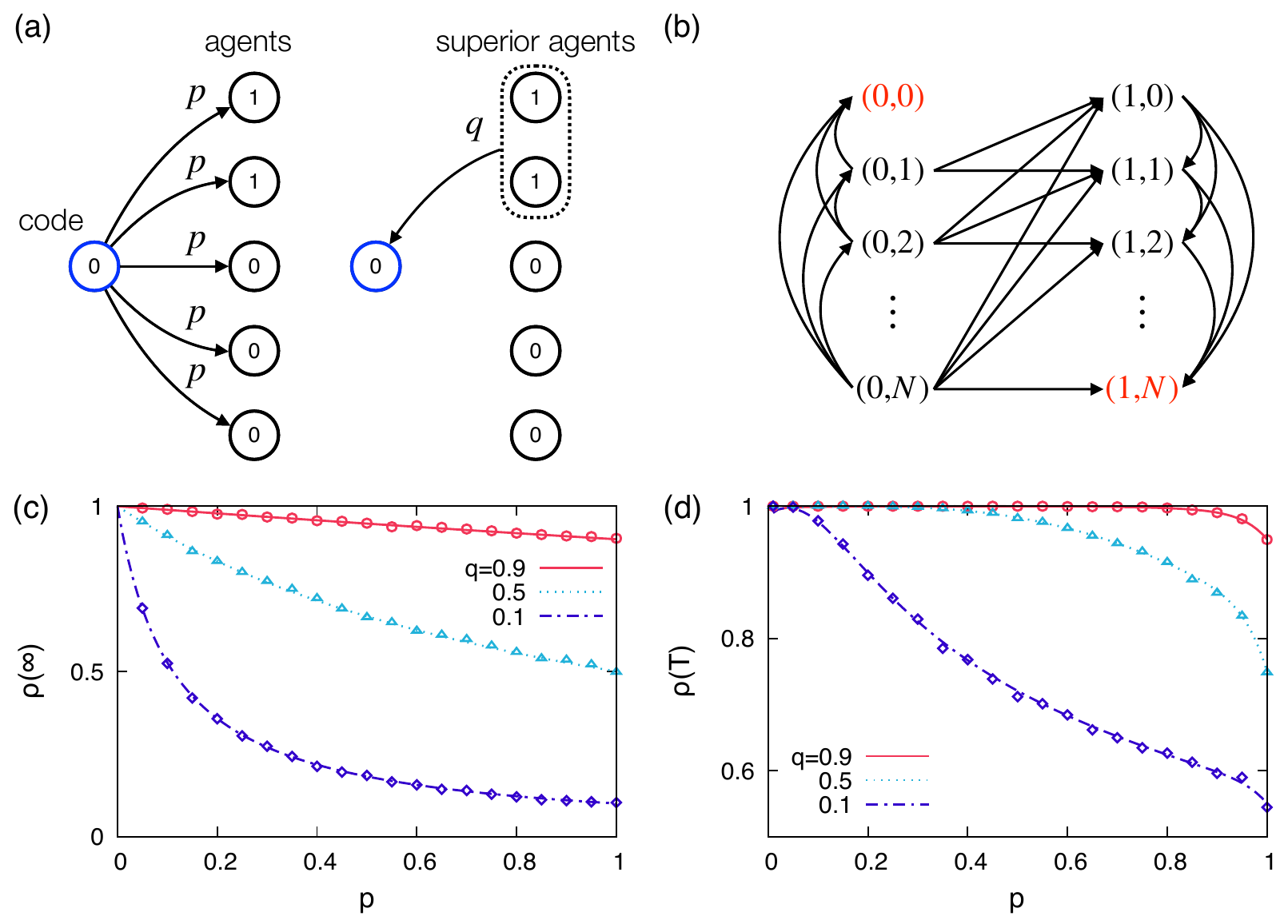}
\caption{(a) Schematic diagram of the homogeneous learning model in a closed system with $r=1$, showing how the code and agents learn from each other with learning probabilities $p$ and $q$ [Eqs.~\eqref{eq:agent_learn}~and~\eqref{eq:code_learn}]. (b) Transition structure between states of the system. Each state is denoted by $(c,n)$, where $c$ is the belief of the code and $n$ is the number of agents whose belief is $1$. Self-loops are not shown for better visualization. Both $(0,0)$ and $(1,N)$ are absorbing states. (c) Analytic solutions of $\rho(\infty)$ in Eq.~\eqref{eq:rho_pq} (lines) with simulation results (symbols) for the case with $N=40$ and an initial condition that $P_{cn}(0)=\delta_{c,0}\delta_{n,1}$. (d) Numerical estimation of $\rho(T)$ using the master equation in Eq.~\eqref{eq:master} (lines) for the case with $N=40$ and an initial condition that $P_{cn}(0)=\frac{1}{2}\delta_{n,N/2}$ for each $c\in\{0,1\}$. Here $T$ is the first time step satisfying $|\rho(T)-\rho(T-1)|<10^{-4}$. The corresponding simulation results are shown with symbols. In panels (c,~d), each symbol was averaged over $10^4$ different runs. Standard errors are omitted as they are smaller than symbols.
}
\label{fig:hom_close}
\end{figure*}

\subsection{Homogeneous learning in a closed system}\label{subsec:hom}

We consider the homogeneous learning model in a closed system. That is, $r(t)=1$ for all $t$, there is no turnover of agents, and every agent has the same learning probability $p$, i.e.,
\begin{align}
    p^{(i)}=p\ \textrm{for}\ i=1,\ldots,N.
    \label{eq:hom_define}
\end{align}
See also Fig.~\ref{fig:hom_close}(a). The state of the system at each time step $t$ can be summarized in terms of the code's belief $c(t)$ and the number of agents whose belief matches the reality, which we denote by $n(t)\in\{0,\ldots,N\}$. Precisely, $n(t)$ is defined as
\begin{align}
    n(t)\equiv \sum_{i=1}^N \delta_{\sigma_i(t),r(t)}.
\end{align}
In our case with $r(t)=1$, one simply has $n(t)=\sum_i \sigma_i(t)$. Then the expected density of agents with the belief matching the reality is given by 
\begin{align}
\rho(t)\equiv \left\langle \frac{n(t)}{N}\right\rangle,
\label{eq:rho_hom_define}
\end{align}
which can be interpreted as the expected belief of a randomly chosen agent, or average individual knowledge~\cite{Blaschke2006Forgotten}.

Depending on the initial belief of the code, two scenarios are possible. Firstly, if $c(0)=1$, the code does not change its belief because it already coincides with the reality, and agents' beliefs will eventually converge to the value of $1$ by Eq.~\eqref{eq:agent_learn}. It implies an absorbing state that the code and all agents share the same value as the reality, which is denoted by $(c,n)=(1,N)$. Secondly, if $c(0)=0$, $n(t)$ will decrease until the code's belief changes to $1$ by Eq.~\eqref{eq:code_learn} as long as there is at least one agent with belief of $1$. Once the code's belief becomes $1$, $n(t)$ will increase to reach the absorbing state $(c,n)=(1,N)$. However, this is not always the case; $n(t)$ may reach $0$ before $c(t)$ changes to $1$, implying that both the code and agents have the belief of $0$ without further dynamics. This indicates another absorbing state $(c,n)=(0,0)$. Figure~\ref{fig:hom_close}(b) shows the transition structure between states with two absorbing states emphasized in red. 

For the analysis, let us denote by $P_{cn}(t)$ the probability that at time step $t$ the code's belief is $c$ and there are exactly $n$ agents with belief of $1$. These probabilities satisfy the normalization condition as
\begin{align}
    \sum_{c=0}^1 \sum_{n=0}^N P_{cn}(t)=1.
\end{align}
They evolve according to the following master equation in discrete time:
\begin{align}
    P_{cn}(t+1)=\sum_{c'n'} W_{c'n'\to cn}P_{c'n'}(t),
    \label{eq:master}
\end{align}
where the transition probabilities read [see Fig.~\ref{fig:hom_close}(b)]
\begin{align}
\begin{split}
    &W_{00\to cn}=\delta_{0,c}\delta_{0,n},\\
    &W_{0n'(\neq 0)\to 0n}=\begin{cases}
    {n'\choose n}p^{n'-n}\bar p^n\bar q & \textrm{if}\ n'\geq n,\\
    0 & \textrm{if}\ n'< n,
    \end{cases}\\
    &W_{0n'(\neq 0)\to 1n}=\begin{cases}
    {n'\choose n}p^{n'-n}\bar p^n q & \textrm{if}\ n'\geq n,\\
    0 & \textrm{if}\ n'< n,
    \end{cases} \label{eq:hom_W} \\
    &W_{1n'\to 0n}=0\ \forall n',n,\\
    &W_{1n'\to 1n}=\begin{cases}    
    {N-n'\choose N-n}p^{n-n'}\bar p^{N-n} & \textrm{if}\ n'\leq n,\\
    0 & \textrm{if}\ n'>n.
    \end{cases}
\end{split}
\end{align}
Here we have used $\bar p\equiv 1-p$ and $\bar q\equiv 1-q$. Calculating Eq.~\eqref{eq:master} recursively with any initial condition $\{P_{cn}(0)\}$, one can in principle obtain $P_{cn}(t)$ for any $c$, $n$, and $t$, hence $\rho(t)$ in Eq.~\eqref{eq:rho_hom_define}, i.e., 
\begin{align}
    \rho(t)=\frac{1}{N}\sum_{c=0}^1\sum_{n=0}^N nP_{cn}(t).
    \label{eq:rho_define}
\end{align}

We focus on steady states of the model. It is obvious that all initial probabilities eventually end up with two absorbing states, i.e., $(c,n)=(0,0)$ and $(1,N)$, implying $P_{00}(\infty)+P_{1N}(\infty)=1$. Thus, the average individual knowledge in Eq.~\eqref{eq:rho_define} reads
\begin{align}
    \rho(\infty)=P_{1N}(\infty)=1-P_{00}(\infty).
    \label{eq:rho_P00}
\end{align}
As the simplest yet nontrivial case, let us consider an initial condition that $P_{cn}(0)=\delta_{c,0}\delta_{n,1}$, namely, $P_{01}(0)=1$ and $P_{cn}(0)=0$ for all other states $(c,n)\neq (0,1)$. Using
\begin{align}
    W_{01\to 01}=\bar p\bar q \equiv\alpha\ \textrm{and}\ W_{01\to 00}=p\bar q\equiv \beta,
\end{align}
the master equations for $P_{01}$ and $P_{00}$ [Eq.~\eqref{eq:master}] are written as follows:
\begin{align}
\begin{split}
    &P_{01}(t+1)=\alpha P_{01}(t),\\
    &P_{00}(t+1)=P_{00}(t)+\beta P_{01}(t).
\end{split}
\end{align}
Master equations for all other states than $P_{01}(t)$ and $P_{00}(t)$ are irrelevant to calculate $P_{00}(\infty)$ in Eq.~\eqref{eq:rho_P00}. Since $P_{01}(t)=\alpha^t$, one obtains
\begin{align}
    P_{00}(t)=\beta(1+\alpha+\ldots+\alpha^{t-1}),
\end{align}
leading to
\begin{align}
    \rho(\infty)=1-\frac{\beta}{1-\alpha}=\frac{q}{p+q-pq}.
    \label{eq:rho_pq}
\end{align}
This solution is not a function of $N$ due to the choice of the initial condition that $P_{cn}(0)=\delta_{c,0}\delta_{n,1}$. 

We observe that $\rho(\infty)$ in Eq.~\eqref{eq:rho_pq} is a decreasing function of $p$ but an increasing function of $q$ [Fig.~\ref{fig:hom_close}(c)], already partly implying the qualitatively similar behavior to the simulation results of the original March model, i.e., Fig.~1 in Ref.~\cite{March1991Exploration}. The larger $q$ leads to the more correct belief of agents about the reality, which is easily understood by considering that $q$ is the learning probability of the code from superior agents. On the other hand, the effect of $p$ on $\rho(\infty)$ is not straightforward to understand. It is because the large value of $p$ speeds up not only the probability flow to the state $(0,0)$ but also to the state $(1,N)$. It means that the large $p$ always helps spread the code's belief to agents whether the code's belief is correct or not. When the code's belief is incorrect, the large $p$ increases the amount of flow to the state $(0,0)$ [Fig.~\ref{fig:hom_close}(a)]. In contrast, when the code's belief is correct, the large $p$ does not increase the amount of flow to the state $(1,N)$, but only speeds up the flow. As we focus on the steady behavior, such an asymmetric role of $p$ leads to the decreasing behavior of $\rho(\infty)$ as a function of $p$. Such a behavior has been interpreted that slow socialization allows for longer exploration, resulting in better organizational learning~\cite{March1991Exploration}.

For general initial conditions, one can estimate $\rho(T)$ for a sufficiently large $T$ by iterating the master equation in Eq.~\eqref{eq:master} for a given initial condition $\{P_{cn}(0)\}$. For a demonstration, we consider a system of $N=40$ agents and the initial condition that $P_{cn}(0)=\frac{1}{2}\delta_{n,N/2}$ for each $c\in\{0,1\}$. We estimate the value of $\rho(T)$ at the first time step $T$ when $|\rho(T)-\rho(T-1)|<10^{-4}$ is satisfied. From the results shown in Fig.~\ref{fig:hom_close}(d), we find that $\rho(T)$ is a decreasing function of $p$ but an increasing function of $q$, showing the same tendency as the solution of $\rho(\infty)$ in Eq.~\eqref{eq:rho_pq} for the simpler initial condition.

These exact and numerical results are supported by agent-based simulations. We perform the simulations of the model using the rules in Eqs.~\eqref{eq:agent_learn}~and~\eqref{eq:code_learn} together with Eq.~\eqref{eq:hom_define} for the system with $N=40$ agents with the mentioned initial conditions. Firstly, the initial condition with $P_{cn}(0)=\delta_{c,0}\delta_{n,1}$ used for the analysis is realized in the simulation such that only one agent has an initial belief of $1$, while all other agents and the code have the belief of $0$. Secondly, as for the initial condition with $P_{cn}(0)=\frac{1}{2}\delta_{n,N/2}$ for each $c\in\{0,1\}$, we set $\sigma_i(0)=1$ for $i=1,\ldots,20$ and $\sigma_i(0)=0$ for the rest of agents, while the value of $c(0)$ is randomly chosen from $\{0,1\}$ with equal probabilities. Eventually every run ends up with one of absorbing states, implying that $n(\infty)=0,N$. For each pair of $p$ and $q$, we take the average of $n(\infty)/N$ over $10^4$ different runs to get the value of $\rho(\infty)$ in Eq.~\eqref{eq:rho_hom_define}. Such averages are shown with symbols in Fig.~\ref{fig:hom_close}(c,~d), which are indeed in good agreement with analytical and numerical solutions, respectively.

\subsection{Heterogeneous learning in a closed system}
\label{subsec:het}

\begin{figure*}[!t]
\centering
\includegraphics[width=0.75\textwidth]{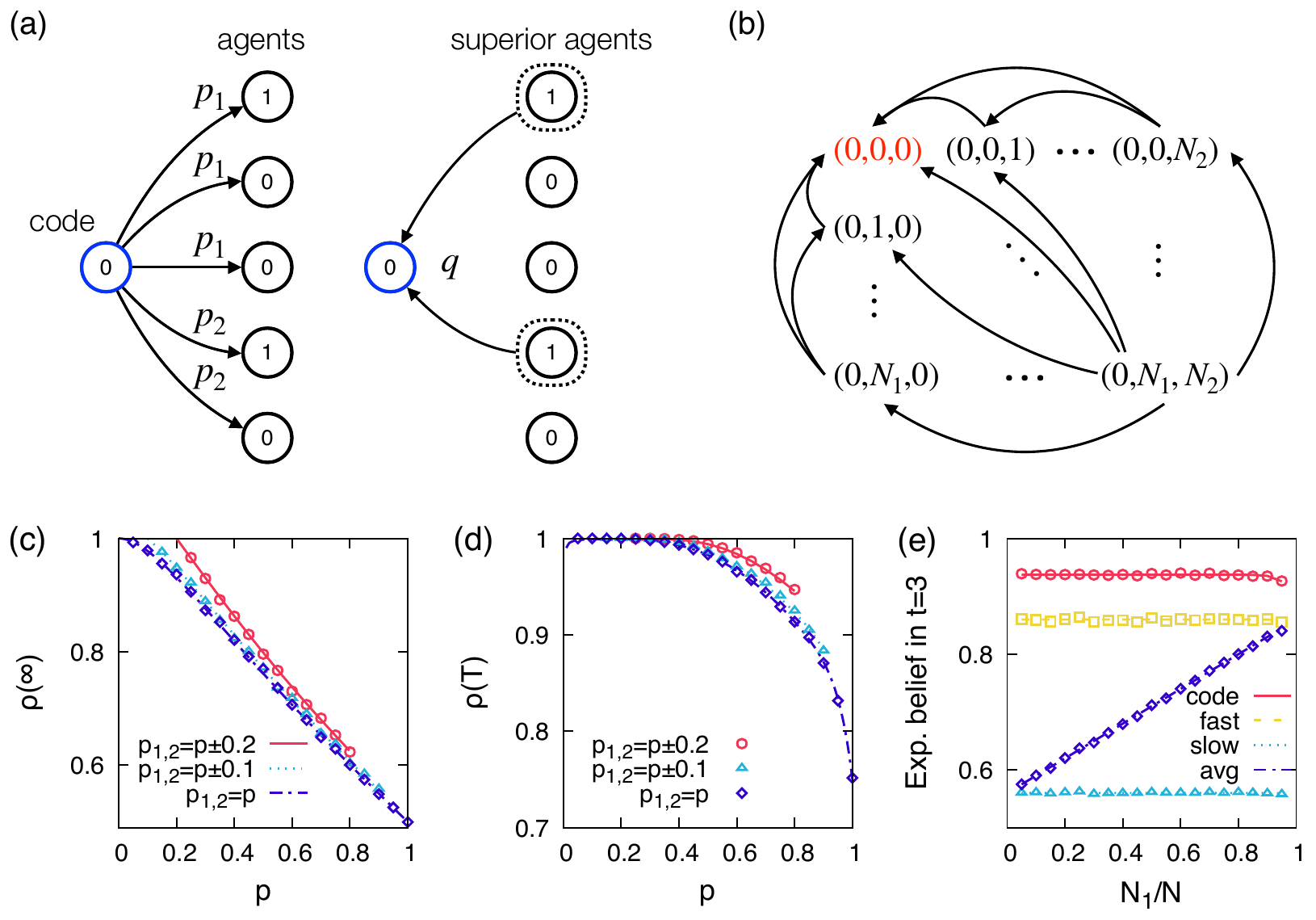}
\caption{(a) Schematic diagram of the heterogeneous learning model in a closed system with $r=1$, showing how the code and agents learn from each other with learning probabilities $p_1$, $p_2$, and $q$. (b) Transition structure between states of the system. Each state is denoted by $(c,n,m)$, where $c$ is the belief of the code and $n$ ($m$) is the number of agents with learning probability $p_1$ ($p_2$) whose belief is $1$. Self-loops and states with $c=1$ are not shown. (c) Analytic solutions of $\rho(\infty)$ in Eq.~\eqref{eq:rho_p1p2q} (lines) for the case with $N=40$ ($N_1=N_2=20$) and an initial condition that $P_{cnm}(0)=\delta_{c,0}\delta_{n,1}\delta_{m,1}$. (d) Numerical estimation of $\rho(T)$ using the master equation in Eq.~\eqref{eq:master2} (lines) for the case with $N=40$ ($N_1=N_2=20$) and an initial condition that $P_{cnm}(0)=\frac{1}{2}\delta_{n,N_1/2}\delta_{m,N_2/2}$ for each $c\in\{0,1\}$. Here $T$ is the first time step satisfying $|\rho(T)-\rho(T-1)|<10^{-4}$. (e) Numerical estimation of expected beliefs of the code (``code'' in the figure), fast-learning agents with $p_1=0.9$ (``fast''), slow-learning agents with $p_2=0.1$ (``slow''), and all agents (``avg'') at $t=3$ using the master equation in Eq.~\eqref{eq:master2} for the case with $N=40$ and $N_1=2,4,6,\ldots,38$ (lines). We use an initial condition that $P_{cnm}(0)=\frac{1}{2}\delta_{n,N_1/2}\delta_{m,N_2/2}$ for each $c\in\{0,1\}$ and for each $N_1$. In panels (c--e), simulation results are shown with symbols, each symbol was averaged over $2\times 10^4$ different runs, and standard errors are omitted as they are smaller than symbols.
}
\label{fig:het_close}
\end{figure*}

Next, we study the heterogeneous version of the model in a closed system with $r=1$ by using two distinct values of learning probability, i.e., by setting
\begin{align}
\begin{split}
    &p^{(i)}=p_1\ \textrm{for}\ i=1,\ldots,N_1,\\
    &p^{(i)}=p_2\ \textrm{for}\ i=N_1+1,\ldots,N,
\end{split}
\end{align}
where $1\leq N_1\leq N-1$ [Fig.~\ref{fig:het_close}(a)]. Agents with the larger (smaller) learning probability among $p_1$ and $p_2$ can be called fast (slow) learners~\cite{March1991Exploration}. The state of the system at each time step $t$ can be summarized in terms of the code's belief $c(t)$, the number of agents with $p_1$ whose belief is $1$, which we denote by $n(t)\in\{0,\ldots,N_1\}$, and the number of agents with $p_2$ whose belief is $1$, which we denote by $m(t)\in\{0,\ldots,N_2\}$. Here $N_2\equiv N-N_1$. Then the expected density of agents with belief of $1$ is given as
\begin{align}
\rho(t)\equiv \left\langle \frac{n(t)+m(t)}{N}\right\rangle,
\label{eq:rho_het_define}
\end{align}
which can also be interpreted as the expected belief of a randomly chosen agent. 

Similarly to the homogeneous version of the model, the master equation reads
\begin{align}
    P_{cnm}(t+1)=\sum_{c'n'm'} W_{c'n'm'\to cnm}P_{c'n'm'}(t),
    \label{eq:master2}
\end{align}
where the transition probabilities are written as
\begin{widetext}
\begin{align}
\begin{split}
    &W_{000\to cnm}=\delta_{0,c}\delta_{0,n}\delta_{0,m},\\
    &W_{0n'm'(\neq 00)\to 0nm}=\begin{cases}
    {n'\choose n}p_1^{n'-n}\bar p_1^n{m'\choose m}p_2^{m'-m}\bar p_2^{m}\bar q & \textrm{if}\ n'\geq n\ \&\ m'\geq m,\\
    0 & \textrm{otherwise},
    \end{cases}\\
    &W_{0n'm'(\neq 00)\to 1nm}=\begin{cases}
    {n'\choose n}p_1^{n'-n}\bar p_1^n{m'\choose m}p_2^{m'-m}\bar p_2^{m} q & \textrm{if}\ n'\geq n\ \&\ m'\geq m,\\
    0 & \textrm{otherwise},
    \end{cases}\\
    &W_{1n'm'\to 0nm}=0\ \forall n',m',n,m,\\
    &W_{1n'm'\to 1nm}=\begin{cases}
    {N_1-n'\choose N_1-n}p_1^{n-n'}\bar p_1^{N_1-n}{N_2-m'\choose N_2-m}p_2^{m-m'}\bar p_2^{N_2-m} & \textrm{if}\ n'\leq n\ \&\ m'\leq m,\\
    0 & \textrm{otherwise}.
    \end{cases}
\end{split}
\end{align}
\end{widetext}
Here we have used $\bar p_1\equiv 1-p_1$ and $\bar p_2\equiv 1-p_2$. It is obvious that there are two absorbing states, i.e., $(0,0,0)$ and $(1, N_1, N_2)$, implying that $P_{000}(\infty)+P_{1N_1N_2}(\infty)=1$. Calculating Eq.~\eqref{eq:master2} recursively with any initial condition $\{P_{cnm}(0)\}$, one can in principle obtain $P_{cnm}(t)$ for any $c$, $n$, $m$, and $t$, hence $\rho(t)$ in Eq.~\eqref{eq:rho_het_define}.

As the simplest yet nontrivial case, let us consider an initial condition that $P_{cnm}(0)=\delta_{c,0}\delta_{n,1}\delta_{m,1}$. Denoting
\begin{align}
\begin{split}
    &\alpha_{i_1i_2}\equiv W_{0i_1i_2\to 0i_1i_2},\\
    &\beta_{i_1i_2,j_1j_2}\equiv W_{0i_1i_2\to 0j_1j_2}\ [(i_1,i_2)\neq (j_1,j_2)],
\end{split}
\end{align}
the master equations for $P_{011}$, $P_{010}$, $P_{001}$, and $P_{000}$ [Eq.~\eqref{eq:master2}] are written as follows:
\begin{align}
    P_{011}(t+1)=&\alpha_{11} P_{011}(t),\nonumber\\
    P_{010}(t+1)=&\alpha_{10} P_{010}(t)+\beta_{11,10} P_{011}(t),\nonumber\\
    P_{001}(t+1)=&\alpha_{01} P_{001}(t)+\beta_{11,01} P_{011}(t),\\
    P_{000}(t+1)=&P_{000}(t)+\beta_{11,00} P_{011}(t)+\beta_{10,00} P_{010}(t)\nonumber\\
    &+\beta_{01,00} P_{001}(t),\nonumber
\end{align}
where $\alpha_{11}=\bar p_1\bar p_2\bar q$,
$\alpha_{10}=\bar p_1\bar q$,
$\alpha_{01}=\bar p_2\bar q$,
$\beta_{11,10}=\bar p_1p_2\bar q$,
$\beta_{11,01}=p_1\bar p_2\bar q$,
$\beta_{11,00}=p_1p_2\bar q$,
$\beta_{10,00}=p_1\bar q$, and
$\beta_{01,00}=p_2\bar q$. After some algebra, one obtains
\begin{align}
    P_{000}(\infty)=&\frac{\beta_{11,00}}{1-\alpha_{11}} +\frac{\beta_{11,10}\beta_{10,00}}{(1-\alpha_{11})(1-\alpha_{10})} \nonumber\\
    &+\frac{\beta_{11,01}\beta_{01,00}}{(1-\alpha_{11})(1-\alpha_{01})},
\end{align}
leading to
\begin{align}
    \rho(\infty)=1-\frac{p_1p_2\bar q}{1-\bar p_1\bar p_2\bar q}\left[1+\frac{\bar p_1\bar q}{1-\bar p_1\bar q}+\frac{\bar p_2\bar q}{1-\bar p_2\bar q}\right].
    \label{eq:rho_p1p2q}
\end{align}
This result is not a function of $N_1$ and $N_2$ due to the choice of the initial condition that $P_{cnm}(0)=\delta_{c,0}\delta_{n,1}\delta_{m,1}$. It is straightforward to prove that setting $p_1=p_2=p$ reduces the solution in Eq.~\eqref{eq:rho_p1p2q} to the solution of the homogeneous model with the initial condition that $P_{cn}(0)=\delta_{c,0}\delta_{n,2}$.

To demonstrate the effect of heterogeneous learning on $\rho(\infty)$ in Eq.~\eqref{eq:rho_p1p2q}, we parameterize $p_1=p+\delta$ and $p_2=p-\delta$ with non-negative $\delta$ and $p\in(\delta, 1-\delta]$. Here $\delta$ controls the degree of heterogeneity of agents. As shown in Fig.~\ref{fig:het_close}(c), the larger $\delta$ leads to the higher values of $\rho(\infty)$ in Eq.~\eqref{eq:rho_p1p2q} for the entire range of $p$, which is consistent with the simulation results of the original March model, i.e., Fig.~2 in Ref.~\cite{March1991Exploration}. Such behaviors can be essentially understood by comparing the transition probability $W_{011\to 000}=p_1p_2\bar q$ in the heterogeneous model to its counterpart $W_{02\to 00}=p^2\bar q$ in the homogeneous model  [Eq.~\eqref{eq:hom_W}] to get
\begin{align}
    \frac{W_{011\to 000}}{W_{02\to 00}}=1-\frac{\delta^2}{p^2}\leq 1.
    \label{eq:ratio}
\end{align}
It implies that for positive $\delta$ the probability flow to the absorbing state $(0,0,0)$ in the heterogeneous model is always smaller than the flow to the absorbing state $(0,0)$ in the homogeneous model, hence the larger $\rho(\infty)$ for the heterogeneous model than for the homogeneous model. We also remark that the ratio in Eq.~\eqref{eq:ratio} gets closer to $1$ for the larger value of $p$, hence the smaller gap between the heterogeneous and homogeneous models. Such an expectation is indeed the case as depicted in Fig.~\ref{fig:het_close}(c).

For general initial conditions, we numerically estimate $\rho(T)$ for a sufficiently large $T$ by iterating the master equation in Eq.~\eqref{eq:master2} for a given initial condition $\{P_{cnm}(0)\}$. For a demonstration, we consider a system of $N=40$ agents ($N_1=N_2=20$) and the initial condition that $P_{cnm}(0)=\frac{1}{2}\delta_{n,N_1/2}\delta_{m,N_2/2}$ for each $c\in\{0,1\}$. We estimate the value of $\rho(T)$ at the first time step $T$ when $|\rho(T)-\rho(T-1)|<10^{-4}$ is satisfied. From the results shown in Fig.~\ref{fig:het_close}(d), we find that $\rho(T)$ has higher values for more heterogeneous systems.

We also perform the agent-based simulations of the heterogeneous model for the system with $N=40$ agents ($N_1=N_2=20$) with the mentioned initial conditions. Firstly, the initial condition with  $P_{cnm}(0)=\delta_{c,0}\delta_{n,1}\delta_{m,1}$ is realized in the simulation such that one agent with $p_1$ and one agent with $p_2$ have an initial belief of $1$, while all other agents as well as the code have the belief of $0$. Secondly, as for the initial condition with $P_{cnm}(0)=\frac{1}{2}\delta_{n,N_1/2}\delta_{m,N_2/2}$ for each $c\in\{0,1\}$, we set $\sigma_i(0)=1$ for $i=1,\ldots,10,21,\ldots,30$ and $\sigma_i(0)=0$ for the rest of agents, while the value of $c(0)$ is randomly chosen from $\{0,1\}$ with equal probabilities. Eventually every run ends up with one of absorbing states, implying that $n(\infty)+m(\infty)=0,N$. For each combination of $p_1$, $p_2$, and $q$, we take the average of $[n(\infty)+m(\infty)]/N$ over $2\times 10^4$ different runs to get the value of $\rho(\infty)$ in Eq.~\eqref{eq:rho_het_define}. Such averages are shown with symbols in Fig.~\ref{fig:het_close}(c,~d), which are indeed in good agreement with analytical and numerical solutions, respectively.

Finally, we note that our setup for heterogeneous agents is different from that in the original March model~\cite{March1991Exploration}. In the original paper, the heterogeneity was controlled by the number of agents having $p_1$, i.e., $N_1$, while learning probabilities were fixed to be $p_1=0.9$ and $p_2=0.1$. We test such original setup using our simplified model both by estimating $\rho(3)$ from the master equations in Eq.~\eqref{eq:master2} and by performing agent-based simulations up to $t=3$. For a system of $N=40$ agents, we consider $N_1=2,4,6,\ldots,38$ with the initial condition that $P_{cnm}(0)=\frac{1}{2}\delta_{n,N_1/2}\delta_{m,N_2/2}$ for each $c\in\{0,1\}$. In addition to the expected belief of all agents in Eq.~\eqref{eq:rho_het_define}, we measure the expected belief of fast-learning agents with $p_1$, that of slow-learning agents with $p_2$, and that of the code. Results from the numerical estimation of master equations and from agent-based simulations are in good agreement with each other as depicted in Fig.~\ref{fig:het_close}(e). These results show the qualitatively same behaviors as in the original March model, i.e., Fig. 3 in Ref.~\cite{March1991Exploration}.

\subsection{Homogeneous learning in an open system}
\label{subsec:hom_turn}

\begin{figure*}[!t]
\centering
\includegraphics[width=0.6\textwidth]{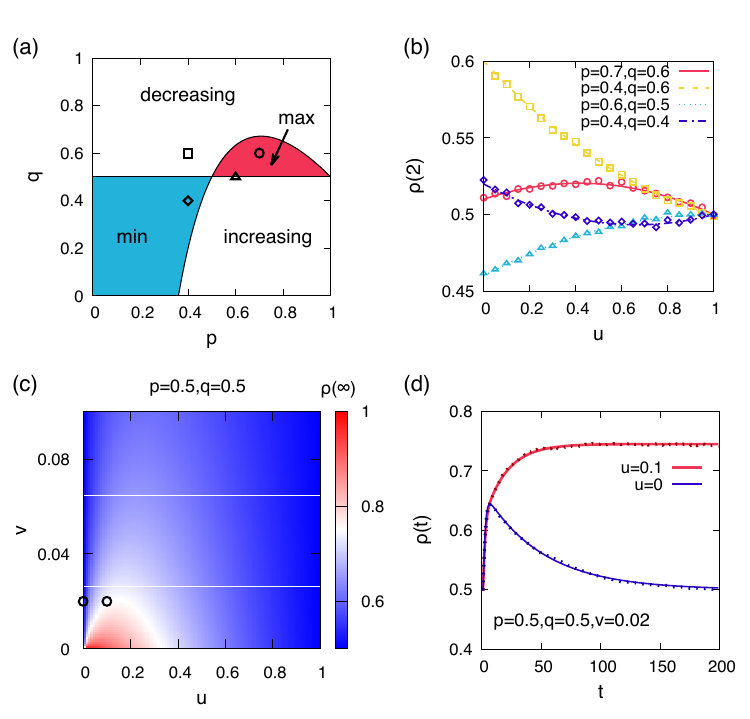}
\caption{(a) Different behaviors of the analytic solution of $\rho(2)$ in Eq.~\eqref{eq:rho_2_pqu} as a function of $u$ depicted in the plane of $(p,q)$. $\rho(2)$ can have either a maximum value for $u\in(0,1)$ (red shade, denoted by ``max'') or a minimum value for $u\in(0,1)$ (tin shade, denoted by ``min''), or it can monotonically increase (``increasing'') or decrease (``decreasing''). Four empty symbols are chosen to demonstrate their different functional forms of $\rho(2)$ in panel (b). (b) Analytic solutions of $\rho(2)$ in Eq.~\eqref{eq:rho_2_pqu} as a function of $u$ for several combinations of $p$ and $q$ (lines) with corresponding simulation results (symbols). We use an initial condition that $P_{c\sigma}(0)=\delta_{c,0}\delta_{\sigma,1}$. (c) Heatmaps of the analytic solution of $\rho(\infty)$ in Eq.~\eqref{eq:rho_pquv} as a function of $u$ and $v$ for the case with $p=q=0.5$. Two empty symbols are chosen to demonstrate different behaviors of $\rho(t)$ in panel (d). (d) Numerical estimation of $\rho(t)$ using the master equation (solid lines) and simulation results (dotted lines) for $u=0$ and $0.1$ when $p=0.5$, $q=0.5$, and $v=0.02$. 
In panels (b,~d), each symbol was averaged over $2\times 10^5$ different runs. Standard errors are omitted as they are smaller than symbols.
}
\label{fig:pquv}
\end{figure*}

We finally study the effects of turnover of agents and turbulence of the external reality on organizational learning. For this, we focus on the simplest yet nontrivial case with $N=1$, indicating that there is only one agent in the system. This agent's belief is denoted by $\sigma(t)$. The case with general $N>1$ can be studied too within our framework.

We first consider the system with turnover of agents only, while $r(t)=1$ for all $t$, namely, $u>0$ and $v=0$. Let us denote by $P_{c\sigma}(t)$ the probability that at time step $t$ the code's belief is $c$, and the agent's belief is $\sigma$. These probabilities satisfy the normalization condition as
\begin{align}
    \sum_{c,\sigma\in\{0,1\}} P_{c\sigma}(t)=1.
\end{align}
They evolve according to the following master equation in discrete time:
\begin{align}
    P_{c\sigma}(t+1)=\sum_{c'\sigma'} W_{c'\sigma' \to c\sigma}P_{c'\sigma'}(t),
    \label{eq:master_hom_open}
\end{align}
where the transition probabilities read using $u'\equiv u/2$ and $\bar u'\equiv 1-u'$ as follows:
\begin{align}
\begin{split}
    &W_{00\to 00}=\bar u',\ W_{00\to 01}=u',\\
    &W_{01\to 00}=(p \bar u'+\bar pu')\bar q,\ W_{01\to 01}=(\bar p \bar u'+pu')\bar q,\\
    &W_{01\to 10}=(p \bar u'+\bar pu') q,\ W_{01\to 11}=(\bar p \bar u'+pu') q, \\
    &W_{10\to 10}=\bar p \bar u'+pu',\ W_{10\to 11}=p \bar u'+\bar pu',\\
    &W_{11\to 10}=u',\ W_{11\to 11}=\bar u'.
\end{split}
\label{eq:transition_csigma}
\end{align}
and all other transition probabilities are zero. Note that due to $u>0$, both $(c,\sigma)=(0,0)$ and $(1,1)$ are no longer absorbing states. 

For the steady state, we derive the analytical solution of $\rho(\infty)$ as
\begin{align}
    \rho(\infty)=\sum_{c\in\{0,1\}} P_{c1}(\infty)=\frac{p+(\frac{1}{2}-p)u}{p+(1-p)u}.
    \label{eq:rho_pqu}
\end{align}
This solution is independent of the initial condition, and it is not a function of $q$ because the change of the code's belief from $0$ to $1$ is irreversible; as long as $q>0$, all initial probabilities end up with states with $c=1$. We also find that $\lim_{u\to 0}\rho(\infty)=1$ and $\rho(\infty)=1/2$ for $u=1$, both of which are irrespective of $p$. That is, $\rho(\infty)$ is a decreasing function of $u$ for $u>0$, whereas for $u=0$ it can have the finite value less than one, e.g., given in Eq.~\eqref{eq:rho_pq}, as long as $p>0$. Thus one can conclude that $\rho(\infty)$ shows an ``increasing'' and then decreasing behavior in the range of $0\leq u\leq 1$. This argument is important to discuss about the optimal turnover of agents that maximizes the effectiveness of organizational learning.

Next, we focus on the transient dynamics instead of the steady state. Starting from the initial condition that $P_{c\sigma}(0)=\delta_{c,0}\delta_{\sigma,1}$, we obtain, e.g., at $t=2$
\begin{align}
    \rho(2)=&(\tfrac{1}{2}-p)(1-p)u^2-(2p^2+pq-\tfrac{7}{2}p+1)u\nonumber \\
    &+(1-p)^2+pq.
    \label{eq:rho_2_pqu}
\end{align}
It turns out that $\rho(2)$ is a quadratic function of $u$, meaning that it can have either a maximum or minimum value for the range of $u\in (0,1)$, or it may be a monotonically increasing or decreasing function of $u$, depending on the choice of $p$ and $q$. Figure~\ref{fig:pquv}(a) summarizes such behaviors in the plane of $(p,q)$, and Fig.~\ref{fig:pquv}(b) depicts $\rho(2)$ as a function of $u$ for several cases of $p$ and $q$. For example, for sufficiently large $q$, $\rho(2)$ is a monotonically decreasing function of $u$ irrespective of $p$. It implies that if the code learns fast from the superior agent, the maximal organizational learning is achieved when there is no turnover. This result can be understood by considering the fact that the turnover introduces randomness or new information from outside of the system. In contrast, if the code learns slowly from the superior agent but the agent learns fast from the code, the maximal organizational learning is achieved for the largest turnover. In such case, without turnover, both the code and agent are likely to be stuck in a suboptimal situation. Thus, the strong turnover may help the system to evade it. Precisely, we find the increasing and then decreasing behavior of $\rho(2)$ for $p=0.7$ and $q=0.6$, and the monotonically decreasing behavior of $\rho(2)$ for $p=0.4$ and $q=0.6$, which are consistent with the results of the original March model, e.g., Fig.~4 in Ref.~\cite{March1991Exploration}.

We now consider the effect of turbulence on the organizational learning in the presence of turnover of agents. For this, we define an extended system consisting of both the system and the reality, whose states can be denoted by $(r,c,\sigma)\in\{0,1\}^3$. Let us denote by $P_{rc\sigma}(t)$ the probability that at time step $t$ the reality is $r$, the code's belief is $c$, and the agent's belief is $\sigma$. These probabilities satisfy the normalization condition as
\begin{align}
    \sum_{r,c,\sigma\in\{0,1\}} P_{rc\sigma}(t)=1.
\end{align}
They evolve according to the following master equation in discrete time:
\begin{align}
    P_{rc\sigma}(t+1)=\sum_{r'c'\sigma'} W_{r'c'\sigma' \to rc\sigma}P_{r'c'\sigma'}(t).
    \label{eq:master_hom_u}
\end{align}
Denoting $v'\equiv v/2$ and $\bar v'\equiv 1-v'$ and using Eq.~\eqref{eq:transition_csigma}, we get the transition probabilities for Eq.~\eqref{eq:master_hom_u} as follows:
\begin{align}
\begin{split}
    &W_{0c'\sigma' \to 0c\sigma}=\overline{W}_{c'\sigma' \to c\sigma}\bar v',\\
    &W_{0c'\sigma' \to 1c\sigma}=\overline{W}_{c'\sigma' \to c\sigma}v',\\
    &W_{1c'\sigma' \to 0c\sigma}=W_{c'\sigma' \to c\sigma}v',\\
    &W_{1c'\sigma' \to 1c\sigma}=W_{c'\sigma' \to c\sigma}\bar v',
\end{split}
\label{eq:transition_temp}
\end{align}
where we have used
\begin{align}
    \overline{W}_{c'\sigma' \to c\sigma}\equiv W_{1-c',1-\sigma' \to 1-c,1-\sigma}.
\end{align}
As $r(t)$ is no longer constant, the average individual knowledge is obtained as
\begin{align}
    \rho(t)=\sum_{r,c,\sigma\in\{0,1\}} \delta_{r,\sigma} P_{rc\sigma}(t).
    \label{eq:rho_pquv_define}
\end{align}

After some algebra, we derive an exact solution of $\rho(\infty)$ for the steady state as follows:
\begin{widetext}
\begin{align}
    \rho(\infty)=& [(p^2q-p^2-\tfrac{5}{2}pq+2p+q-1)u^2v^2
    -(p^2q+p^2-\tfrac{7}{2}pq+2p+\tfrac{3}{2}q-1)u^2v 
    -(p-\tfrac{1}{2})qu^2 
    -(2p^2q-2p^2-\tfrac{9}{2}pq +3p\nonumber \\ &
    +\tfrac{3}{2}q-1)uv^2
    +(2p^2q-2p^2-\tfrac{9}{2}pq+2p+q)uv +pqu
    +(p^2q-p^2-2pq+p+\tfrac{1}{2}q)v^2  
    -p(pq-p-q)v]/[2(p^2q-p^2 \nonumber \\ &
    -2pq+2p+q-1)u^2v^2 
    -(2p^2q-2p^2-5pq+4p+3q-2)u^2v 
    +(1-p)qu^2  
    -(4p^2q-4p^2-8pq+6p+3q-2) \nonumber \\ &
    \times uv^2
    +(4p^2q-4p^2-7pq+4p+2q)uv  
    +pqu  
    +(2p^2q-2p^2-4pq+2p+q)v^2  
    -2p(pq-p-q)v].
    \label{eq:rho_pquv}
\end{align}
\end{widetext}
This analytical solution is depicted as a heatmap in Fig.~\ref{fig:pquv}(c) for the case with $p=q=0.5$. We find that for each value of turbulence $v$ there exists an optimal turnover probability $u^*\in(0,1)$ that maximizes the effectiveness of organizational learning. Such an optimal turnover probability for a given $v$ is obtained as
\begin{align}
    u^*(v)=\frac{v^2+3v-\sqrt{2v(v+3)(3v+1)}}{v^2-3v-2},
\end{align}
which is an increasing function of $v$. It implies that the system is required to have the larger turnover to adapt to the more turbulent external reality. Yet the value of $\rho(\infty)$ with the optimal turnover tends to decrease with $v$ [Fig.~\ref{fig:pquv}(c)].

Finally, we look at the transient dynamics of $\rho(t)$ for different values of $u$ when $p$, $q$, and $v$ are given. We numerically obtain $\rho(t)$ by iterating the master equation in Eq.~\eqref{eq:master_hom_open} using the initial condition that $P_{rc\sigma(0)}=1/8$ for each state. Numerical results are depicted as solid lines in Fig.~\ref{fig:pquv}(d). The agent-based simulations are also performed using the initial condition that each of $r(0)$, $c(0)$, and $\sigma(0)$ is randomly and independently drawn from $\{0,1\}$. Simulation results are shown as dotted lines in Fig.~\ref{fig:pquv}(d), which are in good agreement with numerical results. These results are also qualitatively similar to those in the original March model, e.g., Fig.~5 in Ref.~\cite{March1991Exploration}.

\section{Conclusion}\label{sec:conclusion}

In our work, the celebrated organizational learning model proposed by March~\cite{March1991Exploration} has been simplified, enabling us to explicitly write down the master equation for the dynamics of the model. We have successfully derived exact solutions for the simplest yet nontrivial cases and numerically estimated quantities of interest using the master equations, both results are found to be in good agreement with agent-based simulation results. Our results help rigorously understand not only the simplified model but also the original March model to a large extent.

Our theoretical framework for the simplified March model can be applied to the original March model as well as variants of the March's model that incorporate other relevant factors such as forgetting of the beliefs~\cite{Blaschke2006Forgotten, Miller2016Organizational} and direct interaction and communication between agents in the organization~\cite{Miller2006Adding, Kane2007Information, Kim2009Exploration}. For modeling the interaction structure between agents, various network models might be deployed~\cite{Borgatti2009Network, Barabasi2016Network, Newman2018Networks, Menczer2020First}. In conclusion, we expect to gain deeper insights into the organizational learning using our analytical approach.




\begin{acknowledgments}
H.-H.J. acknowledges financial support by the National Research Foundation of Korea (NRF) grant funded by the Korea government (MSIT) (No. 2022R1A2C1007358).
\end{acknowledgments}




%

\end{document}